# HISTORY BASED COALITION FORMATION IN HEDONIC CONTEXT USING TRUST

Ahmadreza Ghaffarizadeh and Vicki H. Allan

Department of Computer Science, Utah State University, Logan, UT 84341, USA
ghaffarizadeh@aggiemail.usu.edu, vicki.allan@usu.edu

## ABSTRACT

*In this paper we address the problem of coalition formation in hedonic context. Our modelling tries to be as realistic as possible. In previous models, once an agent joins a coalition it would not be able to leave the coalition and join the new one; in this research we made it possible to leave a coalition but put some restrictions to control the behavior of agents. Leaving or staying of an agent in a coalition will affect on the trust of the other agents included in this coalition. Agents will use the trust values in computing the expected utility of coalitions. Three different risk behaviors are introduced for agents that want to initiate a coalition. Using these risk behaviors, some simulations are made and results are analyzed.*

## 1. INTRODUCTION

A multiagent system (MAS) is a system composed of various interacting agents. Based on the type of agents, system can be a simple group or a complicated system doing sophisticated operations. Agents in a MAS can have different types of objectives they are trying to reach. There is not any external decision-maker or centralized processing unit in a MAS; agents work autonomously and communicate with each other to do their tasks [1]. Due to lack of resources or expertise, agents may not be able to reach their goals by themselves, so they are willing to form coalition in order to reach their goals jointly. In most applications of multiagents, agents are considered as cooperative, this means that they form coalition to reach common goals. This research explores the context of *hedonic games*; in this context agents are self-interested and their coalition satisfaction depends on the composition of the coalition they join. Agents only care about their own profit and try to maximize it; so it is not reasonable for these agents to follow actions which maximize the global profit not their own profit. Therefore agents in hedonic games define a preference order over the set of all coalitions they can join and use this preference order to form a satisfactory coalition. Many important problems in multi-agent systems including marriage problem, roommate selection, admission proposals in universities, etc. can be categorized in the context of hedonic games.

The main usage of hedonic games is in the game theory models in Economics; different models have been studied in this domain and some research has been conducted to employ the hedonic games concept in Economics models [2,3,4]. These models concentrate on the analysis of agents' behavior; these research studies do not consider a multiagent framework and usually analyze self-interested agents theoretically. Hedonic games are also frequently studied in the context of Artificial Intelligence; these researches basically deals with finding the core solution in Hedonic games, analyzing stability of core and examining its optimality [5,6,7,8]. Some works have been published about coalition formation under uncertainty [7,9,10,11], however these works are not in the context of Hedonic games and can't be adopted.

An interesting research study has been published by Genin and Aknine [12]; this research mainly deals with autonomous agents that are self-interested, a protocol is defined for coalition formation process and some strategies are suggested for proposal acceptance and proposal selection mechanisms. The current research is an extension of this study and tries to consider more realistic cases by introducing new behaviors, parameters and mechanisms.

## 2. PROBLEM DESCRIPTION

### 2.1. Preliminaries

*Definition 1:* Agent $a$ is an autonomous entity who can sense the environment and make a sequence of actions upon to these sensations in order to reach a predefined objective.

*Definition 2:* A multiagent system $M$ is a group of $n$ agents $M=\{a_1, a_2, \ldots, a_n\}$ that are joined to do a/some task(s). These agents may or may not share common resources.

*Definition 3:* Coalition $c$ is a nonempty subset of $M$. Let $C$ denotes the set of all coalitions, then $C_i$ denotes all coalitions which agent $i$ belongs to; $C_i = \{c \subset M \mid a_i \in c\}$.

*Definition 4:* A Hedonic game $G$ is a preference order $\succ_i$ defined over sets of agents; $(M, \succ_i)$; this preference order is transitive and asymmetric.

*Definition 5:* A solution for a Hedonic game is a partition from set of all agents M, $\{c_1, c_2, \ldots, c_k\}$ where $\sum_{j=1}^{k} c_j = M$ and $\forall i \neq j,\ c_i \cap c_j = \emptyset$.

### 2.2. Protocol

We do not have any central decision maker and agents in this research decide for themselves via negotiation with other agents. During time each agent may receive many proposals for forming coalition; it also sends some proposals to other agents asking about their willingness to form coalition. We use a simple two-phase protocol: an agreement phase and commitment phase. The protocol works as follow:

1- An initiator selects one of its possible coalitions according to its preference operator (it will be discussed in later sections); suppose that this coalition includes $k$ agents (1 initiator and $k$-1 other agents).
2- Initiator sends the proposal of the selected coalition to all $k$-1 agents.
3- Agents respond to initiator with yes or no about their interest in joining this coalition.
4- If all agents agreed with the proposed coalition, initiator sends a confirmation request to all agents and wants them to commit to the coalition, otherwise it sends a cancellation message.

If all agents commit to the coalition, initiator informs all members that coalition is formed otherwise it sends a cancellation message.

## 3. PROPOSAL ACCEPTANCE STRATEGY

### 3.1. Proposal Acceptance

Agents in this research are rational; since proposal acceptance does not have any cost, and there is always a probability that the current coalitions get broken; agents accept all proposals they receive except for the proposals in which they have a utility less than when they are alone. Each proposal has a limited waiting time, i.e. agents should respond by that time; otherwise the response will be considered as rejection. Three types of agents are considered in this research:

1- Early responders who responds as soon as possible.
2- Lazy responders who waits to the end of deadline.
3- Random responders who have a random behavior in responding to the proposals.

### 3.2. Proposal Confirmation

On confirming proposals, whenever an agent feels that a proposal for a coalition is better than current coalition, it can leave current one and join the proposed coalition. Following are some considerations in confirming proposals:

1- Each coalition has an obligatory staying time $T_c$, it means that if an agent enrolled in a coalition, it needs to be in the coalition for at least $T_c$ unit of time. If it wants to leave coalition sooner, it should pay the penalty $P_c$. Earned money from penalty will be shared between all agents of the coalition equally.
2- Each agent has a level of honesty $H_i$; since leaving a coalition is unpleasant from viewpoint of other members, the expected utility of leaving current coalition and joining the other one should be great enough to tempt agent $i$ to join a new coalition; as an example if utility of current coalition is 1000; an agent with honesty level of 0.1 will be tempted by a coalition with utility more than 1100 and an agent with level of 0.4 will be tempted by a coalition with utility more than 1400. There is another factor that discourages agents to leave one coalition: Bad Reputation Coefficient (*BRC*). To decide on leaving current coalition and joining other coalition, agents act in a greedy way: they only consider the next timestep. Totally, each agent evaluates the following criterion for the next step and if it holds, then it leaves the current coalition and joins the new one.

    Expected utility of current coalition*(1+$H_i$) < Expected utility of proposed coalition*(1- *BRC*)

3- There would be an enrollment fee for each solicited agent, $Cost_{Enroll}$ ; the initiator will receive a part of this fee as a reward for forming the coalition which benefits all.

Since confirmation doesn't mean that the confirmed proposal by this agent will certainly form a coalition, the agent may need to confirm some proposals before joining one of them; but there is a restriction that an agent at each time step can only confirm a limited number of proposals.

### 3.3. Trust between Agents

Each agent $a_i$ has a record for all other agents that shows its trust in them to not leave the coalition. As an example, $tr_{ij}$ shows the trust $a_i$ has in the agent $a_j$ to stay in the next step in the current coalition; from mathematical point of view, $tr_{ij}$ is the probability that $a_j$ won't leave the group in the next time step. All trust values are initialized by 0.5. In each time step all agents update their trust values for other included agents in the coalition; suppose that agent $a_i$ in time step $t$ wants to update its trust records for agent $a_j$ that is in the same coalition as $a_i$ is; two cases may happen:

1- If $a_j$ stayed in the current coalition in step $t$-1, then $a_i$ increases the value $tr_{ij}$ by the special trust reward $r$.
2- If $a_j$ has left the current coalition in step $t$-1, then $a_i$ decreases the value $tr_{ij}$ by the special betray punishment $pu$.

The value for $pu$ should be bigger than $r$ to reflect that other agents hate the betrayer agent that breaks the coalition. In this research $r$ is 0.01 and $pu$ is considered as 0.05.

### 3.4. Expected Utility Evaluation

Agent $i$ uses the following formula to calculate the expected utility of current coalition $C_{cur}$ for the next step as:

$$ExpectedUtility_i(C_{cur}) = Utility_i(C_{cur}) \times \prod_{a_j \in C_{cur}} tr_{ij}$$

To compute the expected value of a proposed coalition $C$, agent $i$ uses the following formula:

$$ExpectedUtility_i(C) = Utility(C) - (Penalty_{leaving} + Cost_{enroll}) \times stepCoeff$$

*stepCoeff* is a coefficient that lets the agent to consider the likelihood of being more than one time step in the proposed coalition; so the *stepCoeff* is less than one.

### 4. PROPOSAL SELECTION STRATEGY

We consider a cost $Cost_{Com}$ for each proposal the initiator sends. The initiator should consider this cost to avoid sending bad proposals, i.e. the proposals that has a little chance of acceptance by all agents. Initiator needs a measure to calculate the likelihood of formation of a candidate coalition; so it calculates degree of interest of solicited agents based on the similarity with previous proposals.

Similar to work done by Genin and Aknine[12] and based on the Bayesian reinforcement learning model proposed by Chalkiadakis and Boutiller[11], the normalized distance between two coalitions is used in this research to estimate their similarity. Suppose that all agents store all sent and received proposals. Therefore if agent $i$ finds a coalition that is similar to previously formed coalition, it can hope that this candidate coalition will be formed too. The normalized distance between two coalitions $c_1$ and $c_2$ is defined as:

$$d(c_1, c_2) = \frac{1}{n} |(c_1 \cup c_2) \setminus (c_1 \cap c_2)|$$

For each agent $a_j$ in candidate coalition $c$, initiator calculates $d_j^{rcv}(c)$, $d_j^{acc}(c)$ and $d_j^{ref}(c)$ as the distance between $c$ and all the proposals received from $a_j$, all the accepted proposals sent to $a_j$ and all the refused proposals sent to $a_j$ accordingly. Using these three values, initiator computes the following degree of interest for $a_j$:

$$\delta_j(c) = \min(d_j^{rcv}(c), d_j^{acc}(c)) - d_j^{ref}(c)$$

$\delta_j(c)$ less than zero shows that $a_j$ may be interested in $c$ and when $\delta_j(c)$ is greater than zero it shows that $a_j$ is not. Since a coalition will be formed if and only if all agents are interested, so for analyzing the likelihood of formation of a coalition, the degree of most unwilling agent should be International Journal of Artificial Intelligence & Applications (IJAIA), Vol. 4, No. 4, July 2013considered. If we map the willingness value to the range [0, 1], we can use this value as the probability of formation of a coalition ($P_{formation}$).

At each step, agents can send a limited number of proposals. If agents were banned to leave a formed coalition, initiator could decide on its preference order and evaluate proposals one by one; but in this research, agents can leave coalitions. So it is possible that an agent, who is not interested in a proposal in time step $t$, accepts it at time step $t+k$. Also when the number of agents is too large, it is not possible to have a preference order due to cost of memory and computation. Based on these facts, agents in this project do not have a preference order; instead they have a preference operator based on their risk attitude:

1- Risk seeking: agents in this category weigh the utility of coalition more than the possibility of formation. They risk proposing the coalitions with high utility that may have low possibility of formation. They use following measure to choose between the proposal they can send:
$$Preference(C) = Utility(C)^\alpha \times P_{formation}(C) \quad (\alpha > 1)$$

2- Risk-averse: these agents prefer to choose the proposal that has the higher chance to be accepted. Following equation is the measure which agents in this category use to discriminate different proposals:
$$Preference(C) = Utility(C) \times P_{formation}(C)^\beta \quad (\beta < 1)$$

3- Risk-neutral: these agents consider expected value and chance of acceptance in the same weight; using this measure they decide which proposal to send:
$$Preference(C) = Utility(C) \times P_{formation}(C)$$

At each time step, an initiator creates a set that contains a) a fixed number of random coalitions b) some randomly selected proposals from the previously sent proposals and c) some randomly selected proposals from previously received proposals. Using this set, initiator selects the best proposals it can send using its preference operator. Agents do not send the selected proposal if its expected utility is less than a threshold; this threshold is set to $2*Cost_{Com}$ in our work.

## 5. EXPERIMENTS AND RESULTS

We conducted 4 different simulations and repeated each simulation 20 times. There are 20 agents in each simulation running for 100 time steps. Bad Reputation Coefficient is set to 0.15 and Honesty level for each agent is taken from the range [0, 0.35]. Expiration time for accepting and confirmation is set to 3 steps. To calculate the utility of coalitions, each agent considers the direct interaction with other agents and pairwise interaction of each two other agents in the coalition; each agent has a table of estimation of interaction for each two agents (including itself). It finds all permutation of size 2 of agents and look up for the value in the estimation table, then sums all these value to compute the utility of a given coalition. Each entry in the table is randomly chosen from the range [-100, 100].

For the first simulation, all agents are risk seeking; for the second simulation agents are risk averse and in the third simulation all the agents are risk neutral. We have considered the equal number of each type of agents in the fourth simulation. Gained utility for agents with similar honesty levels are averaged for simulations and are depicted in Figure 1. Note that honesty levels are divided to 0.05 ranges and each agent is mapped to the closest level in averaging. Results show that risk-seeking and risk-neutral agents have similar utility behavior over the changes in honesty factor and have the highest utility with the honesty factor [0.15, 0.20].

Second set of results show the average distribution of initiator, solicited and alone agents per step in each simulation. Note that axes are scaled to show the distribution clearly. Figure 2 shows the

International Journal of Artificial Intelligence & Applications (IJAIA), Vol. 4, No. 4, July 2013

distribution of alone agents, it can be seen that in the simulation with risk-seeking agents the number of alone agents are less than two other simulations. Figure 3 depicts distribution of solicited agents that have similar distributions in all three simulations. Distribution of initiator agents are depicted in Figure 4; since each coalition has only one initiator, Figure 4 shows the number of formed coalitions as well. As results show, number of formed coalitions in simulation with risk-neutral agents is less than two other simulations. The average duration of coalitions for risk-seeking agents is 8 steps, for risk-averse agents is 12 and for risk neutral agents is 17 steps.

## 6. CONCLUSION

Humans are self-interested and always (with some exceptions and restrictions) follow their preferences in their interaction with others. That is why context of "hedonic games," that includes self-interested agents, has been received a significant attention. In this paper we addressed the problem of coalition formation in the context of hedonic games. We modelled a realistic case where agents at each time step can decide to stay in a coalition or leave. As in reality, when an agent leaves a group, other agents lose their trust in the left agent. Agents have an honesty factor; the more this honesty factor, the more reliable our agents are. Agents use trust values to estimate the possibility of continuing with the coalition they currently are in. We analyzed the gained utility of agents regards to their risk behavior in proposing coalitions to

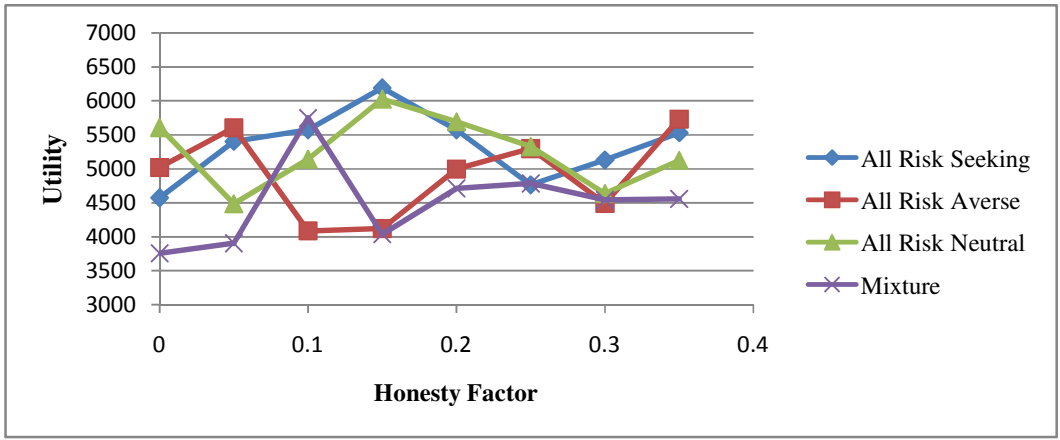

Figure 1: Utility versus Honesty factor regards to agent types in 4 different simulations.

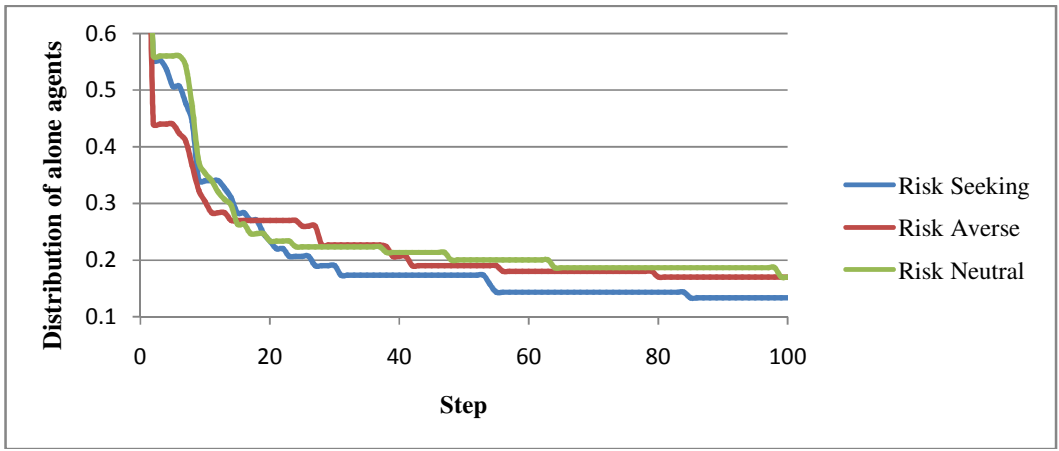

Figure 2. Distribution of alone agents in 100 time steps for agents with different risk attitudes.

International Journal of Artificial Intelligence & Applications (IJAIA), Vol. 4, No. 4, July 2013

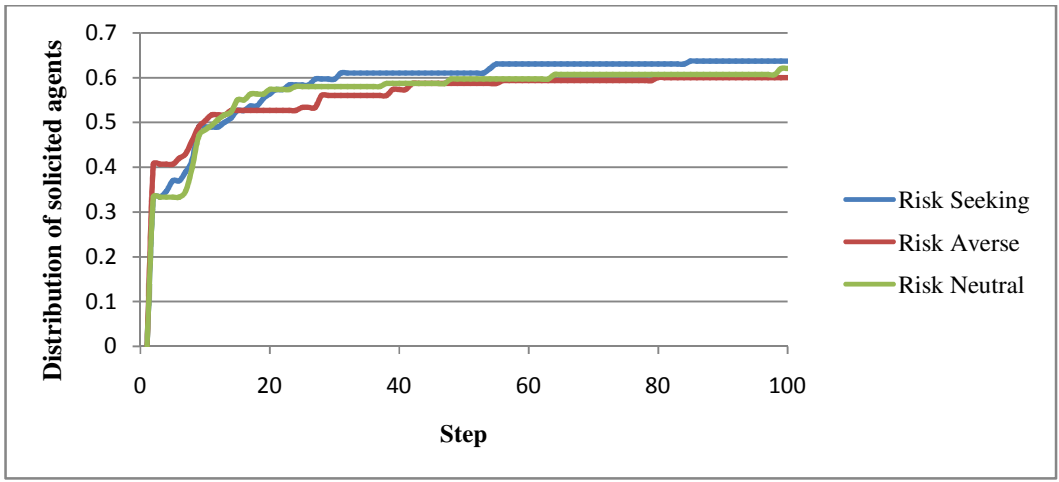

Figure 3. Distribution of solicited agents in 100 time steps for agents with different risk attitudes.

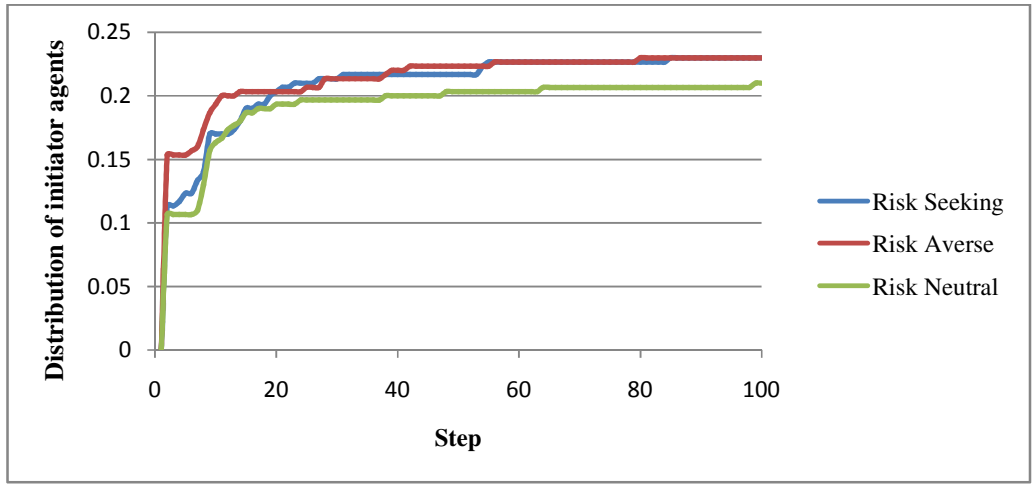

Figure 4. Distribution of initiator agents in 100 time steps for agents with different risk attitudes.

 other agents. Results show that different levels of honesty can lead to different utility values for distinct risk attitudes.

## 7. FUTURE WORK

Some ideas for future work are as follows: one may think about considering the level of friendship between agents; an agent may accept a proposal from a friend while it doesn't accept a similar proposal with a higher utility from a stranger. As in real world, some people like to be manager, some agents may enjoy of being initiator; it would be interesting to consider a factor to show the willingness of agents about being an initiator or being solicited. Implementing this behavior can be done by introducing a value α in the range $[0,1]$ showing the willingness of each agent to be an initiator. An agent with $α=0$ never propose coalitions to others and an agent with $α=1$ never accept any proposals from others; other values will affect the decision of being an initiator or not. As another future work, one may employ history to estimate duration of candidate coalitions to use in the coalition utility evaluation.

## REFERENCES

[1] K. Ahmadi and V. H. Allan, "Efficient Self Adapting Agent Organizations," in Proceedings of the 5th International Conference on Agents and Artificial Intelligence, ICAART 2013, Barcelona, Spain.

[2] A. Bogomolnaia, M. O. Jackson, "The stability of hedonic coalition structures," Games and Economic Behavior, vol. 38, no. 2, pp. 201-230, Feb. 2002.

[3] S. Banerjee, H. Konishi, and T. Sonmez, "Core in a simple coalition formation game," Social Choice and Welfare, vol. 18, no. 1, pp. 135-153, Jan. 2001.

[4] E. Diamantoudi and L. Xue, "Farsighted stability in hedonic games," Social Choice and Welfare, vol. 21, no. 1, pp. 39-61, Aug. 2003.

[5] S. Sung and D. Dimitrov, "On core membership testing for hedonic coalition formation games," Operations Research Letters, vol. 35, no. 2, pp. 155-158, Mar. 2007.

[6] T. Genin and S. Aknine, "Constraining Self-Interested agents to guarantee pareto optimality in multiagent coalition formation problem," vol. 2, pp. 369-372, Aug. 2011.

[7] F. Bloch and E. Diamantoudi, "Noncooperative formation of coalitions in hedonic games," International Journal of Game Theory, vol. 40, no. 2, pp. 263-280, May 2011.

[8] H. Aziz, F. Brandt and H. G. Seedig, "Stable partitions in additively separable hedonic games," Proc. of AAMAS'11, pp. 183-190, 2011

[9] S. Kraus, O. Shehory, and G. Taase, "Coalition formation with uncertain heterogeneous information," in Proceedings of the second international joint conference on Autonomous agents and multiagent systems, ser. AAMAS '03. New York, NY, USA: ACM, 2003, pp. 1-8.

[10] K. Westwood and V. H. Allan, "Heuristics for dealing with a shrinking pie in agent coalition formation," pp. 537-546, Dec. 2006.

[11] G. Chalkiadakis and C. Boutilier, "Bayesian reinforcement learning for coalition formation under uncertainty," in Proceedings of the Third International Joint Conference on Autonomous Agents and Multiagent Systems - Volume 3, ser. AAMAS Washington, DC, USA: IEEE Computer Society, 2004, pp. 1090–1097

[12] T. Genin and S. Aknine, "Coalition formation strategies for multiagent hedonic games," Tools with Artificial Intelligence, IEEE International Conference on, vol. 1, pp. 465-472, 2010.

## Authors

Ahmadreza Ghaffarizadeh is a research assistant and PhD candidate in Computer Science department at Utah State University. His research interests include computational biology, artificial intelligence and evolutionary algorithms.

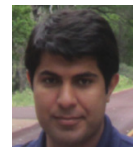

Vicki H Allan completed her PhD in Computer Science at Colorado State University in 1986. She completed a master's degree in computer science and a master's degree in mathematics from Utah State University. She is currently an associate professor in the Computer Science department at Utah State University where she teaches courses in multiagent systems, programming languages, data structures, and algorithms. Her research is supported by NSF grants 0812039 and 0829563.

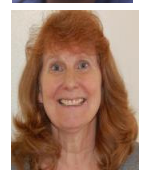